\documentstyle[aps]{revtex}
\input{psfig}

\begin{document}

\textheight 8.8in
\textwidth 6.5in
\topmargin -.25in
\oddsidemargin -.25in
\evensidemargin 0in
\baselineskip 14pt
\def\hm{\ \rm {\it h}^{-1} Mpc}

\title{Tracking Extended Quintessence}

\author{Carlo Baccigalupi$^{1}$\footnote{bacci@sissa.it}, 
Sabino Matarrese $^{2,3}$\footnote{matarrese@pd.infn.it},
Francesca Perrotta$^{1}$\footnote{perrotta@sissa.it}}
\address{$^{1}$ SISSA/ISAS, Via Beirut 4, 34014 Trieste, Italy;\\
$^{2}$ Dipartimento di Fisica `Galileo Galilei', Universit\'a 
di Padova,\\ 
and INFN, Sezione di Padova, Via Marzolo 8, 35131 Padova, Italy;\\
$^{3}$ Max-Planck-Institut f\"ur Astrophysik,
Karl-Schwarzschild-Strasse 1, D-85748 Garching, Germany.} 

\baselineskip 10pt
\maketitle
\begin{abstract}
We investigate the cosmological role of a Tracking Field 
$\phi$ in Extended Quintessence scenarios, where the 
dynamical vacuum energy driving the acceleration of the universe 
today possesses an explicit coupling with the Ricci scalar, $R$, 
of the form $F(\phi )R/2$, where $F(\phi )$ mimics 
General Relativity today, $F(\phi_{0})=1/8\pi G$. 
We analyse explicit non-minimally coupled (NMC) models 
where $F(\phi) =1/8\pi G+\xi(\phi^{2}-\phi_{0}^{2})$, with $\xi$ 
is the coupling constant and $\phi_{0}$ is the Q value today.
Tracker solutions for these NMC models, 
with inverse power-law potentials, possess an initial enhancement of
the scalar field dynamics, named $R$-boost, caused by the effective 
potential generated by the Ricci scalar in the Klein-Gordon equation.  
During this phase the field performs a ``gravitational" 
slow rolling until the true potential becomes important. 
We give accurate analytic formulas describing the $R$-boost, 
showing that the Quintessence energy in this phase 
scales with the redshift $z$ as $(1+z)^{2}$. 
When the $R$-boost ends, the field trajectory matches  
the tracker solution in minimally coupled theories.  
We compute perturbations in these Tracking Extended Quintessence 
models, by integrating the full set of equations for the evolution of 
linear fluctuations in scalar-tensor theories of gravity, and assuming 
Gaussian scale-invariant initial perturbations. 
The Integrated Sachs Wolfe (ISW) effect on the 
Cosmic Microwave Background (CMB) angular spectrum causes a change 
$\delta C_{\ell}/C_{\ell}\simeq 2[1-8\pi G F(\phi_{dec})]$ 
at $\ell\lesssim 10$, where $dec$ stands for decoupling. 
Similarly, the CMB acoustic peaks multipoles shift compared to ordinary 
tracking Quintessence models by roughly an amount 
$\delta\ell /\ell\simeq [8\pi G F(\phi_{dec})-1]/8$. 
The turnover wavenumber $k_{turn}$ 
in the matter power spectrum shifts by an amount 
$\delta k_{turn}/k_{turn}\simeq [1-8\pi G F(\phi_{eq})]/2$, 
where $eq$ stands for matter-radiation equivalence. 
All these corrections may assume positive as well as negative 
values, depending on the sign of the NMC parameter $\xi$. 
We show that the above effects can be as large as $10 - 30\%$ 
with respect to equivalent cosmological constant and ordinary 
tracking Quintessence models, respecting all the existing 
experimental constraints on scalar-tensor theories of gravity.  
These results demonstrate that the playground where 
the next decade data will have their impact includes the  
nature of the dark energy in the Universe, as well as the structure
of the theory of gravity. 
\end{abstract}

\section{Introduction}

Following the experimental evidence in favor of an accelerating 
universe \cite{Perlm,Riess,Garna}, the cosmological role of 
a minimally-coupled scalar field, considered as a "Quintessence" 
component (which we call here Q or $\phi$, equivalently), 
providing the dominant contribution to the
energy density of the Universe today in the form of dynamical vacuum 
energy, has been widely analyzed in the recent past \cite{Stain1} -\cite{CF}. 
The main feature of this dynamical vacuum energy component, that could
also allow to distinguish it from a cosmological constant, is its 
time-dependence and the wide range of possibilities for its equation
of state. Moreover, by the principle of general covariance  
this time varying scalar field should also develop spatial
perturbations.   

In a  number of recent papers the available range 
of couplings between Q and other physical components has been explored. 
The possible coupling between a Quintessence field and light matter fields
has been investigated in \cite{carroll} and is subject to restrictions
coming from the observational constraints on the time variation of the
constants of nature. Also, some recent work explores the 
cosmological consequences 
of a coupling between Quintessence and matter fields \cite{ame0}. 
The possible coupling between Q and the Ricci scalar $R$ was the
subject of our recent analysis \cite{PBM}, where the observable impact
of such a coupling has been throughly investigated. Within this  
`Extended Quintessence' (EQ) model two sets of theories were studied: 
Induced Gravity (IG, originally proposed in \cite{Zee}) and 
Non-Minimal Coupling (NMC, see \cite{Birrel} for an extensive 
overview). In both models the term describing the coupling between 
the Ricci scalar $R$ and the Quintessence field $\phi$ in the 
Lagrangian has the form $F(\phi )R/2$, where $F(\phi )=\xi \phi^{2}$; 
the difference between the two models is that in the NMC model 
the Lagrangian contains also the ordinary gravitational 
term $R/16\pi G$, that is instead absent in IG. Both models belong to
the general class of scalar-tensor theories of gravity. 
The effects on Cosmic Microwave Background (CMB) 
anisotropies, as well as on Large Scale Structure (LSS), 
have been described and numerically computed for several 
values of $\xi >0$; they turned out to be similar for IG and NMC,
although for different values of the coupling constant.  
The main results of our previous study were a modification of the
low redshift dynamics, enhancing the Integrated Sachs Wolfe 
effect (ISW), as well as a shift of the CMB acoustic peaks 
and matter power spectrum turnover because of the dynamics 
of the Hubble length. 
Predicting in detail the shape and the position of the acoustic peaks 
in a given cosmological model is of course important 
in view of the formidable observations of the next decade 
\cite{CMBFUTURE}, but it is also of great importance at present, 
because of the strong evidence in favor of the existence 
of subdegree acoustic oscillations in the latest CMB data 
\cite{Boom,Max}.  

As detailed in our previous paper \cite{PBM}, we named our models 
`Extended Quintessence', in analogy with the `extended inflation'
models \cite{exte}, proposed in the late eighties to rescue the
original idea of inflation based on a first-order phase transition.  
In those models a Jordan-Brans-Dicke (JBD) scalar field \cite{JBD} was 
added to the action to solve the `graceful exit' problem of `old 
inflation'. 
The idea of using a non-minimally coupled scalar
field as a decaying cosmological
constant dates back to a paper by Dolgov \cite{dolgov} in 1983, 
which however radically differs from the present class of models 
by the absence of a true potential energy for the scalar field. 
The background dynamical properties of a Quintessence field 
coupled to the Ricci curvature have also been recently studied in
\cite{Chiba}, \cite{Ame} and \cite{Uzan}, \cite{DeRitis} 
where, in particular, the existence and stability of 
cosmological scaling solutions was analyzed. Models in 
scalar-tensor theories of gravity involving scalar matter 
couplings have also been studied \cite{Pietro}. 

Let us come now to the subject of the present work. The existence of
a considerable amount of vacuum energy in the Universe, as
observations seem to imply, brings together two conceptual problems. 
The first regards the magnitude of this energy, namely the fact that 
we are right now in the phase in which the vacuum energy is 
starting to dominate over matter. 
The second, which we name ``fine-tuning" problem, which arises merely 
from the fact that if the vacuum energy is constant, like in the 
`standard' cosmological constant scenario, then, at the beginning of 
the radiation era the Q energy density should have been vanishingly small 
compared with both radiation and matter. 

Important recent work demonstrated that the Quintessence 
scenario can avoid such fine tunings, while a cosmological constant 
fitting the data would be affected by both problems. 
Indeed, in the Quintessence scenario one can select a subclass of
models which admit "tracking solutions" \cite{track}: 
following early work by Ratra and Peebles \cite{Ratra}, 
it was shown how the observed amount of scalar field energy density
today can be reached starting from a very wide set of initial
conditions, covering tens of orders of magnitude. In particular, the 
Quintessence could have been initially at the level of the ordinary
matter, thus being likely one of the products of the decay 
of the vacuum energy responsible for the inflation era. 

Here we face the extension of tracking phenomenology 
in EQ; we find important results, both for background 
dynamics and perturbations, and we name this scenario 
Tracking Extended Quintessence (TEQ). 
We consider a NMC model for the coupling between Q and $R$. 
We first analyse the background evolution, accurately 
studying the Quintessence dynamics in the radiation dominated 
(RDE) and matter dominated (MDE) eras. 
Then, we compute the evolution of cosmological perturbations 
pointing out important observable effects on CMB and LSS. 

The work is organized as follows: in Sec. II we recall 
the relevant equations, and we refer to previous works 
\cite{PBM,CF} for a complete presentation of the dynamical system 
for the background and for the perturbations in non-minimally 
coupled scalar field cosmologies; 
Sec. III contains the analysis of tracking solutions; 
Secs. IV and V contain the description and a discussion of 
the effects on CMB and LSS, respectively; Sec. VI discusses
how these effects change if we vary the form of the 
Q-field potential.Finally, Sec. VII contains 
a summary of our results and some concluding remarks. 

\section{Cosmology in scalar-tensor theories of 
gravity}

In this Section we review the relevant equations in 
scalar-tensor theories of gravity, for the background 
evolution and linear perturbations. For more details, 
we refer to previous works \cite{PBM,CF}. 

Following \cite{HW1}, scalar-tensor theories of gravity are 
represented by the action 
\begin{equation}
\label{action} 
S=\int d^4 x \sqrt{-g} \left[ {1 \over 2} f(\phi, R) -
{1 \over 2} \omega ( \phi) \phi^{; \mu} \phi_{; \mu} -V( \phi)
+ L_{fluid}\right]\ ,
\end{equation} 
where $R$ is the Ricci scalar, 
and $L_{fluid}$ includes all components except for 
$\phi$, that we suppose coupled only with gravity. 

The models for $f(\phi ,R)$ that we investigate here are simply a
product of a function of $\phi$ and $R$; 
also we assume a standard form for the kinetic part: 
\begin{equation}
\label{model}
f(\phi, R)=F(\phi) R\ \ ,\ \ \omega (\phi )=1\ . 
\end{equation}

As in \cite{PBM}, we assume the cosmological metric has the 
flat Friedmann-Robertson-Walker (FRW) form: 
\begin{equation}
\label{ds2}
ds^2=a^2 [-d\tau ^2 + (\delta_{i j }+h_{i j })dx^i dx^j ]  \ ,  
\end{equation}
where $a(\tau )$ is the scale factor, $\tau$ is the 
conformal time and $h_{ij}$ represents the perturbations 
in the synchronous gauge, to be discussed later. 
Our units are $c \equiv 1$, and we adopt the signature $(-,+,+,+)$.

In the next subsections we describe the evolution equations 
for the relevant quantities, for the background and linear 
perturbations. 

\subsection{Background}

The background FRW equations read 
\begin{equation} 
\label{Friedmann}
{\cal H}^{2}=\left({\dot{a}\over a}\right)^{2}= {1 \over 3F} \left(
a^2 {\rho}_{fluid} +{1 \over 2} \dot{\phi}^2 + 
a^2 V - 3 {\cal H} \dot{F} \right) \ ,
\end{equation}
\begin{equation}
\label{Friedmann2} 
\dot{\cal H}={\cal H}^2- {1 \over 2F} \left(
a^2({\rho}_{fluid}+p_{fluid})
+ \dot{\phi}^2 + \ddot{F} -2 {\cal H} \dot{F}  \right)\ ,
\end{equation}
where the dot denotes 
differentiation  with respect to the conformal time $\tau$; 
note that in our framework $1/F$ today must be $8\pi G$. 

The Klein-Gordon equation reads 
\begin{equation}
\label{KG}
\ddot{\phi}+2{\cal H} \dot{\phi}= {a^2 \over 2} 
 F_{\phi} R -  a^2 V_{\phi}  \;,
\end{equation}
where the subscript $\phi$ denotes differentiation w.r.t. $\phi$. 
Furthermore, the continuity equations for the 
individual fluid components are not directly affected by the
changes in the gravitational field equation, 
and for the $i-$th component we have 
\begin{equation}
\label{continuity}
\dot{\rho}_i = -3 {\cal H} ({\rho}_i + p_i)\ .
\end{equation}

Let us give now useful expressions for the Ricci scalar 
in scalar-tensor theories: 
\begin{equation}
\label{Ricci}
R=-{1 \over F} \left[ 
-\rho_{fluid}+3p_{fluid}+ 
{ \dot{\phi}^2 \over a^2 } - 4 V +3\left( {\ddot{F} \over
a^2}+ 2 { {\cal H} \dot{F} \over a^2} \right)
 \right]\ .
\end{equation}
This can also be written as a function of the Hubble parameter 
and its derivatives: 
\begin{equation}
R={6 \over a^2} ( {\dot{\cal H}}^2+  {\cal H}^2)\ .
\label{riccih}
\end{equation}
As discussed in more detail in \cite{PBM}, let us recall the two 
experimental constraints that these models have to satisfy. 

One comes from the time variation of the gravitational 
constant in scalar-tensor gravity theories 
\footnote{We consider $1/8\pi F$ as effective time dependent 
gravitational constant. We checked that, for the small 
values of $\xi$ considered here, our definition 
practically coincides with the more general one given in 
\cite{DAMOUR}.}
\begin{equation}
\left|{G_{t}\over G}\right|=-{F_{t}\over F}\ ,
\label{gtg}
\end{equation}
where the subscript $t$ denotes differentiation w.r.t. the 
cosmic time $t$; $G_{t}/G$ is bounded by local laboratory 
and solar system experiments \cite{Gbounds} to be 
\begin{equation}
{G_{t}\over G}\le 10^{-11}\ {\rm per\ year}\ .
\label{sse}
\end{equation}
Another independent experimental constraint comes 
from the effects induced on photon trajectories \cite{omega500}: 
\begin{equation}
\omega_{JBD} \equiv {F_{0}\over F_{\phi 0}^{2}}\ge 500\ ,
\label{jbd}
\end{equation}
where $F_{\phi 0}$ is the derivative of $F$ w.r.t. $\phi$ 
calculated at the present time. 
The latter bound does not depend on the particular cosmological 
trajectory at hand, since in contrast to (\ref{sse}), it 
does not involve time derivatives, but only the present 
value of the field.  

Let us define now the form of $F(\phi )$ that we consider here. 
We analyse non-minimally coupled scalar field models, 
defined as in \cite{PBM}. 
The term multiplying the curvature scalar $R$ is made of a 
dominant contribution, which is a constant, plus one 
depending on $\phi$: 
\begin{equation}
F(\phi )\equiv {1\over 8\pi G}+{\tilde F}(\phi )-
{\tilde F}(\phi_{0})\ ,
\label{nmcstructure}
\end{equation}
where 
\begin{equation}
{\widetilde F}(\phi) = \xi\phi^2\ .
\label{NMC}
\end{equation}
As it is evident from the previous equation, the present value of $F(\phi )$  
has been setted in such a way that the gravitational constant reduces to the observed
one; it is not our aim here to discuss how this result can be dynamically achieved. 
More general scalar-tensor theories in which Einstein gravity is an attractor have 
been discussed in the frame of Quintessence scenarios (see e.g. \cite{Pietro}). 
 
The second constraint
(\ref{jbd}) turns out to be the dominant one for our models; as is very easy to verify, 
it gives rise to the following condition: 
\begin{equation}
\xi\lesssim {10^{-2}\over 2 \sqrt{G}
\phi_{0}}\ .
\label{conjbdnmc}
\end{equation}
The only remaining quantity to specify in our model is 
the potential for the field $\phi$, responsible for 
the the dominant part of the vacuum energy today. 
We adopt here inverse power-law potentials, as originally suggested by
Ratra and Peebles \cite{Ratra}, and found in some 
phenomenological supersymmetry breaking models 
\cite{track,SM}, although cosine \cite{CDF} and exponential 
\cite{Stain2} potentials have also been proposed in the literature. 
We take 
\begin{equation}   
V(\phi )={M^{4+\alpha}\over\phi^{\alpha}}\ ,
\label{v}
\end{equation}
where the value of $\alpha >0$ will be specified later and 
the mass-scale $M$ is fixed by the level of 
energy contribution today from the Quintessence. 

Let us conclude this subsection by defining the 
quantities describing the Q energy density and pressure. 
By considering the kinetic and potential terms, we 
generalize in a straightforward way 
the definitions in ordinary General Relativity:
\begin{equation}
\label{rhoq}
\rho_\phi={1\over 2a^{2}}\dot{\phi}^{2}+V(\phi )\ .
\end{equation}
Similarly, the Quintessence pressure assumes the form 
\begin{equation}
\label{pq}
p_\phi={1 \over 2a^{2}}\dot{\phi}^{2}-V(\phi )\ .
\end{equation}
However, as it is evident from equation (\ref{Friedmann}), 
there are other terms in the right-hand side that could 
be included into the Q energy density and pressure, coming 
directly from the coupling with $R$. Thus we define here the 
generalized Q energy density and pressure, 
that include all the NMC terms:
\begin{equation}
\label{rhophi}
{\tilde\rho}_{\phi}=
{1\over 2a^{2}}\dot{\phi}^{2}+V(\phi ) - {3{\cal H}\dot{F}\over a^{2}}\ .
\end{equation}
\begin{equation}
\label{pphi}
{\tilde p}_{\phi}=
{1\over 2a^{2}}\dot{\phi}^{2}-V(\phi ) + {\ddot{F}\over a^{2}}+
{{\cal H}\dot{F}\over a^{2}}\ .
\end{equation}
We stress that, as we'll see below, in our model the NMC terms are 
negligible at the present time, so that, from the point of view of 
observational tests at low redshifts, there is no appreciable 
difference between the two definitions of energy density and pressure. 

In the next Section we will see how $\rho_\phi,p_\phi$
may differ substantially from ${\tilde\rho}_{\phi},{\tilde p}_{\phi}$, 
while in general they become nearly identical in the MDE. 

\subsection{Perturbations}

As in \cite{PBM,CF}, we adopt here the treatment of the 
scalar perturbations based on the formalism developed in \cite{HW1} to 
describe the evolution of perturbations in the synchronous gauge. 
The perturbing tensor in Eq.(\ref{ds2}) is Fourier transformed and 
the generic component at wavevector ${\bf k}$ may be written as 
\begin{equation}
\label{hij}
h_{i j}({\bf k},\tau )= 
{\bf {\hat{k}_i  \hat{k}_j}}  h( {\bf k}, \tau)
+ ( {\bf {\hat{k}_i  \hat{k}_j}}- {1 \over 3} \delta_{ i j} )
6 \eta  ( {\bf k}, \tau)\ ,
\end{equation} 
where $h$ denotes the trace of $h_{ij}$ and $\eta$ 
represents the traceless component; we omit the 
arguments $({\bf k},\tau )$ in the following. 

The perturbed Einstein equations read
\begin{eqnarray}
\label{t00}
k^{2}\eta -{1\over 2}{\cal H}\dot{h} &=& -{a^2 \delta \rho 
\over 2}  \ ,\\
\label{ti0}
k^{2}\dot{\eta} &=& { a^{2} (p+\rho)\theta \over 2}  \ ,\\
\label{tii}
\ddot{h}+2{\cal H}\dot{h}-2k^{2}\eta &=& -3 a^{2} \delta p \
,\\
\label{tij}
\ddot{h}+6\ddot{\eta}+2{\cal H}(\dot{h}+6\dot{\eta})-2k^{2}\eta &=&
-3 a^{2}(p+\rho)\sigma\ .
\end{eqnarray}

The perturbed density, pressure, velocity and shear 
in the synchronous gauge assume the form
\begin{equation}
\label{deltarho}
\delta \rho = {1 \over F } \left[
\delta {\rho}_{fluid} + {\dot{\phi} \delta\dot{\phi} \over a^2} 
+{1 \over 2} (-F_{\phi} R + 2V_{\phi}) \delta \phi
-3{ {\cal H} \delta \dot{F} \over a^2} -
\left(  { \rho + 3 p \over 2 } + {k^2 \over a^2} \right) \delta F +
{\dot{F}\dot{h} \over 6a^{2}}      \right] \ ,
\end{equation}
\begin{equation}
\label{deltap}
\delta p = {1 \over F } \left[ \delta p_{fluid} +
{\dot{\phi} \delta\dot{\phi} \over a^2}
+ {1 \over 2} (F_{\phi} R - 2V_{\phi}) \delta \phi
+{\delta \ddot{F}\over a^{2}} + { {\cal H} \delta \dot{F} \over a^2} +
\left(  {p- \rho  \over 2 } + {2k^2 \over 3a^2} \right) \delta F 
-{1 \over 9 } {\dot {F} \dot{h}  \over a^2}       \right]\ ,
\end{equation}
\begin{equation}
\label{theta}
(p+ \rho) \theta = {(p_{fluid}+ \rho_{fluid} ) \theta_{fluid}  
\over F }
- {k^2 \over a^2 } \left( 
{ - \dot{\phi} \delta \phi - \delta\dot{F } +{\cal H} \delta
F \over F} \right)\ ,
\end{equation}
\begin{equation}
\label{shear}
(p+ \rho) \sigma=  {(p_{fluid}+ \rho_{fluid} ) \sigma_{fluid}  \over
F }
+ {2k^2 \over 3a^2 F}  
 \left(    \delta F + 3 { \dot{F} \over k^2} ( \dot{\eta} + {\dot{h} 
\over 6} )  \right)  \ , 
\end{equation}
where everything labeled with $fluid$ contains contributions from all the species 
but Quintessence, and has the form as in ordinary General Relativity, obeying the
perturbed continuity equations; we refer to \cite{CF} for a detailed definition of
these terms. There remains the perturbed Klein-Gordon equation: 
\begin{equation}
\label{KGpert}
\delta \ddot{\phi} + 2 {\cal H} \delta\dot{\phi} + \left[ k^2 
+ a^2 {\left( {-F_{\phi} R +2V_{\phi} \over 2} \right) }_{\phi}
\right] \delta \phi=  {\dot{\phi} \dot{h} \over 6 } + {a^2 \over 2}
F_\phi \delta R \ . 
\end{equation}
Note the presence of the Ricci curvature scalar $R$ in the left-hand
side, as well as its perturbation 
$\delta R$ in the right-hand side: the latter term is not trivial, 
since it contains $\delta\ddot{\phi}$ 
(as it can be easily seen by using eqs. (\ref{tii},\ref{deltap}) and 
(\ref{deltaR}) below; from \cite{HW1} we see that 
in the synchronous gauge it is in fact 
\begin{equation}
\label{deltaR}
\delta R={1\over 3a^{2}}\left(\ddot{h}-3{\cal H}\dot{h}+
2k^{2}\eta\right)\ , 
\end{equation}
so that the Klein Gordon equation, using Eq.(\ref{tii}), 
assumes the complicated form below: 
$$
\delta\ddot{\phi}\left(1-{1\over 2}{F_{\phi}^{2}\over F}\right)+ 
\left[ 2 {\cal H} - {F_{\phi}\over 2 F}\left(\dot{\phi}+
2F_{\phi\phi}\dot{\phi}+{\cal H}F_{\phi}\right)
\right]\delta\dot{\phi} + 
$$
$$
+\left[ k^2 + a^2 {\left( {-F_{\phi}R + 
2V_{\phi} \over 2}\right)}_{\phi}
-{F_{\phi}\over 2 F} \left({1\over 2}a^{2}F_{\phi}R
-a^{2}V_{\phi}+F_{\phi\phi\phi}\dot{\phi}^{2}+
F_{\phi\phi}\ddot{\phi}+{\cal H}F_{\phi\phi}\ddot{\phi}\right)+
\right] \delta \phi +
$$
$$
+F_{\phi}\left({a^{2}p-a^{2}\rho\over 2}+{2\over 3}k^{2}\right)
\delta \phi= 
$$
\begin{equation}
\label{KGpertbetter}
={\dot{\phi} \dot{h} \over 6 } - {F_{\phi}\over 6}
\left({\cal H}\dot{h}-10k^{2}\eta\right)+
{F_{\phi}\over 2 F}a^{2}\delta p_{fluid}-
{F_{\phi}^{2}\over 18 F}\dot{\phi}\dot{h}\ . 
\end{equation}
This set of differential equations can be integrated once 
initial conditions on the metric and fluid perturbations are given; 
in this work we adopt adiabatic initial conditions 
for the various components as well as Gaussian and scale-invariant 
initial perturbations (see \cite{PBM,CF}). 

\section{Tracking Extended Quintessence}

Let us now turn to the study of tracking solutions in our 
model. As we already mentioned in the Introduction, 
the existence of this kind of trajectories was first
pointed out by Ratra and Peebles \cite{Ratra}. 
In \cite{track}, it was realized that they could be used 
in Quintessence scenarios to solve 
the fine-tuning problem on the initial conditions; this is in 
favor of Quintessence compared with the ordinary cosmological 
constant models, which remains affected by this problem. 

For what concerns the background parameters, 
we require the present closure density of Quintessence to be
$\Omega_{\phi}=0.7$, with Cold Dark Matter at $\Omega_{CDM}=0.253$, 
three families of massless neutrinos, baryon content 
$\Omega_{b}=0.047$ and Hubble constant $H_{0}=100h_{100}$ 
Km/sec/Mpc (or $h_{100}=0.65$ thus keeping $\Omega_{b}h_{100}^{2}=0.020$). 
In section V we will explore the dependence 
of our results on the potential index $\alpha$. Here 
we adopt $\alpha =2$. The case $\alpha=1$ was studied in
\cite{PBM}, where however, we did not explore the properties of 
tracking solutions. 
As we will discuss in Section VI, however, the dynamical relevance of 
tracking solutions, becomes more and more relevant, 
as the potential slope increases, so, our main results of \cite{PBM} 
are not essentially modified. 

In Figure \ref{f1} we plot the evolution of the 
energy density of the cosmic fluid components (radiation, matter 
and Quintessence) as a function of redshift, in several models that we are 
going to describe now. 
The plotted curves are $\rho_{r}$, $\rho_{m}$ and 
$\rho_\phi$. 
The high, dotted lines represent radiation and matter, 
which do possess the well known scalings  
$\rho_{r}\sim 1/a^{4}$, $\rho_{m}\sim 1/a^{3}$. Equivalence 
occurs roughly at $1+z_{eq}\simeq 5\times 10^{3}$. 
The other curves represent tracker solutions for the Quintessence 
energy density $\rho_{\phi}$. We will describe in detail  
each curve in the next subsections; let us give here some general 
considerations. For each curve, the constant $M$ in Eq.(\ref{v}) 
has been chosen to produce the required $\Omega_{\phi}$ 
today. The NMC dimensionless coupling constant is taken as  
\begin{equation}
\label{betaxi}
|\xi| =1.5 \times 10^{-2}\ ,
\end{equation}
so as to satisfy the experimental constraints (\ref{jbd}), because 
in all the cases shown in Fig.\ref{f1} the present value of 
$\phi$ is $\phi_{0}\simeq 0.35M_{P}$, where $M_{P}=1/\sqrt{G}$. 
Also we study both positive and negative values of $\xi$. 
$\phi_{0}$ is reached starting from initial conditions 
$\phi_{beg}\ll\phi_{0}$ and $\dot{\phi}_{beg}$ 
which can vary by several tens of orders of magnitude 
thanks to the capability of tracker solutions to remove any fine 
tuning of the initial conditions. In fact, it can be immediately noted 
from the curves in Fig.{\ref{f1}} that, 
despite the huge range of initial values of the energy 
density, the Q component is going to dominate today at the 
chosen level. 
The final aspect  we wish to point out before going 
to a detailed analysis of the dynamics in MDE and RDE, is that 
we plotted $\rho_\phi$ in the figure, which is always 
positive, as it is evident from its definition (\ref{rhoq}). 
This is not true for ${\tilde\rho}_{\phi}$ defined in (\ref{rhophi}), 
because the NMC terms may be negative, and in fact 
they overcome the kinetic and potential energy, 
as we will see below; this feature was first found and discussed 
in \cite{PBM}. 

\subsection{TEQ-trajectories in the Radiation Dominated Era: 
$R$-boost}

To understand what happens, let us study the Klein-Gordon 
equation during the Radiation Dominated Era (RDE). 
The first feature to note is the behavior of the Ricci scalar $R$. 
The dominant fluid is radiation, so that $a\sim\tau$, 
and this could yield the wrong idea that $R=0$ from 
Eq.(\ref{riccih}). Actually this is not true as it 
is evident from Eq.(\ref{Ricci}): the first two terms 
$\rho_{fluid}-3p_{fluid}$ take zero contribution from radiation, but
there is a residual contribution $\rho_{m}$, from the subdominant matter
component. 
Thus, there is a divergence $R\propto 1/a^{3}$
as $a\rightarrow 0$, that holds 
until at least one particle species remains non-relativistic. 
Also, we will see in a moment that the other terms in 
Eq.(\ref{Ricci}), as well as the little dynamics 
implied by the overall factor $1/F$ do not change this
argument; in conclusion the behavior of $R$ in the RDE is 
\begin{equation}
\label{RicciRDE}
R\simeq {8\pi G\rho_{m0}\over a^{3}}=
{3H_{0}^{2}(\Omega_{CDM}+\Omega_{b})\over a^{3}}
\ \ {\rm for}\ a\rightarrow 0\ .
\end{equation}
This implies that on the RHS of the 
Klein Gordon equation (\ref{KG}) the term multiplying $R$ diverges 
as $1/a$ and has the same sign of $\xi$ since $\phi$ 
is assumed to be always positive. This generates a ``gravitational" 
effective potential causing an enhancement of the 
dynamics of $\phi$ at early times, 
that we name ``$R$-boost''. The field accelerates 
until the friction term $2{\cal H}\dot{\phi}$ 
reaches a value that is comparable 
to the $R$-boost term on the RHS. 
After that, Q enters a phase of slow roll 
driven by the friction and the $R$-boost terms only: 
\begin{equation}
\label{KGR-boost}
2{\cal H}\dot{\phi}\simeq {a^{2}RF_{\phi}\over 2}\ .
\end{equation}
The slow roll holds until the true potential 
energy from $V$ becomes important in the Klein 
Gordon equation. 
This dynamics is manifest in Fig.\ref{f2}, where 
the absolute values of the four terms in the Klein Gordon 
equation are plotted. The thin solid line is the potential 
term, that is subdominant until $1+z\lesssim 1000$. 
The heavy dashed line is the friction term and 
the solid heavy one is the $R$-boost term. 
In a very short time the friction grows from zero 
(the initial Q velocity is here taken to be zero) until 
it reaches the $R$-boost term, setting the onset of the slow rolling
regime. The thin dashed line is the Q acceleration, $\ddot{\phi}$, 
which is positive and decreasing initially; then, it 
becomes negative (deceleration) at the cusp corresponding to  
$1+z\sim 10^{7}$, and again positive at 
$1+z\sim 10^{3}$ when the potential 
energy becomes important and the tracker behavior 
in the MDE starts. 

\begin{figure} 
\centerline{
\psfig{file=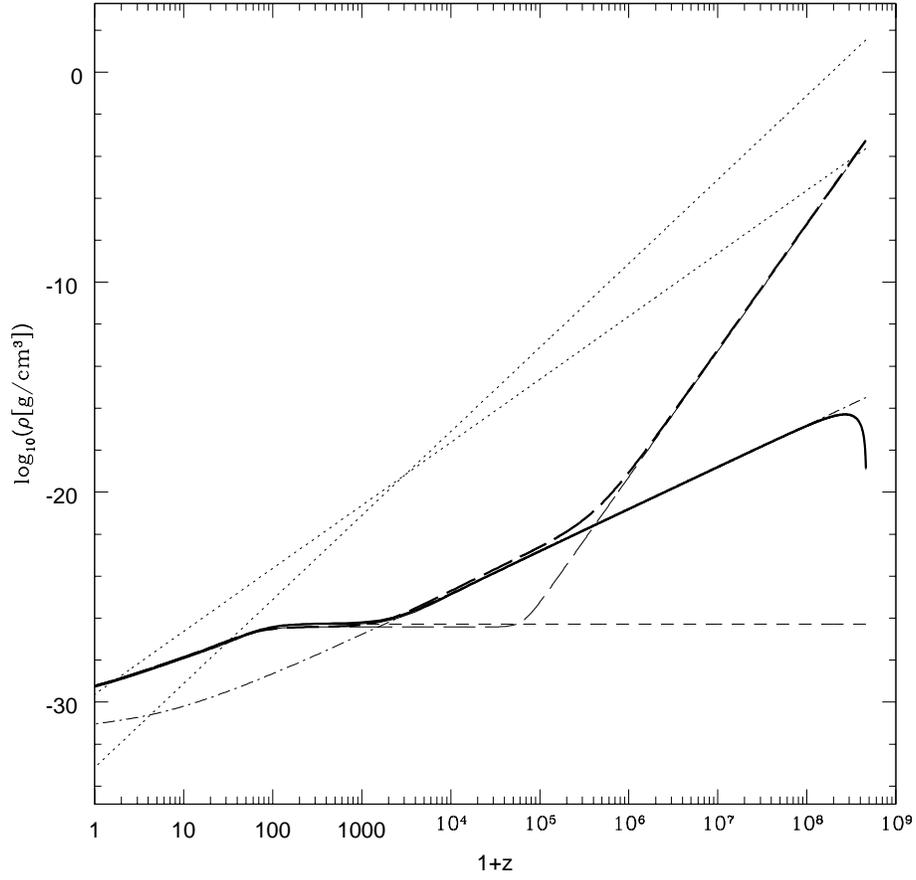,height=5.in,width=5.in}
}
\caption{Matter, radiation, Quintessence energy 
densities as functions of redshift. 
Dotted lines: matter and radiation; 
heavy solid: Tracking Extended Quintessence; 
thin dashed: tracking 
minimally coupled Quintessence; 
heavy dashed: initially highly energetic 
Tracking Extended Quintessence; thin long-dashed 
initially highly energetic tracking Quintessence; 
dot-dashed: $R$-boost analytic.}
\label{f1}
\end{figure}

\begin{figure} 
\centerline{
\psfig{file=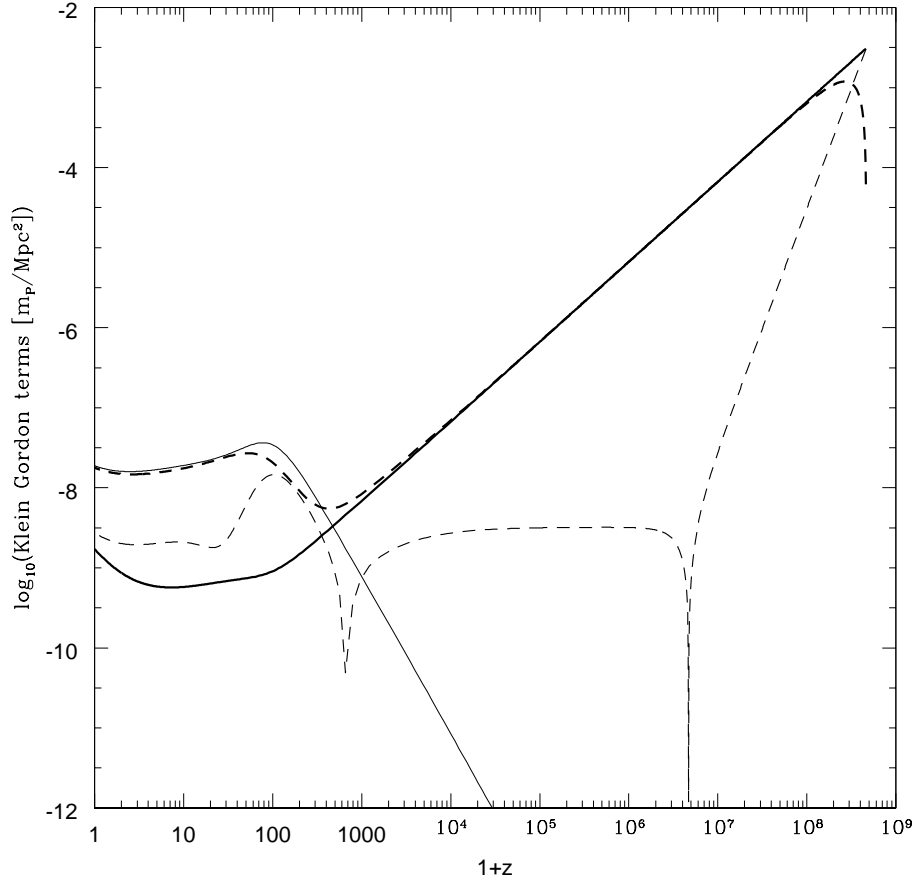,height=5.in,width=5.in}
}
\caption{Absolute value of the Klein Gordon equation terms: 
heavy solid: gravitational effective potential; 
heavy dashed: friction; solid: potential; 
dashed: acceleration.}
\label{f2}
\end{figure}

It is quite simple to write an accurate analytic form 
for the solution during 
the $R$-boost. By using Eqs.(\ref{RicciRDE},\ref{KGR-boost}), 
and the behavior of $a$ in the RDE, $a\simeq 
\sqrt{8\pi G\rho_{r}^{0}/3}\cdot\tau$, we easily get 
\begin{equation}
\label{phiRDE}
\phi =\phi_{beg}\exp{[{\cal C}(\tau -\tau_{beg})]}\ ,
\end{equation}
where $beg$ stands for the time when initial conditions are given, and  
\begin{equation}
\label{c}
{\cal C}={3H_{0}^{2}(\Omega_{CDM}+\Omega_{b})
\over 2\sqrt{8\pi G\rho_{r}^{0}/3}}\xi\ .
\end{equation}
With our choice of parameters, ${\cal C}\simeq 7\cdot 10^{-3}\xi$ 
Mpc$^{-1}$, which implies that the exponent in (\ref{phiRDE}) 
remains much smaller than one at all relevant times. Therefore, 
\begin{equation}
\label{phiRDEapprox}
\phi\simeq\phi_{beg}[1+{\cal C}(\tau -\tau_{beg})]\ .
\end{equation}
The behavior expressed by Eqs.(\ref{phiRDE}, \ref{phiRDEapprox}) 
corresponds to the dotted dashed line in Fig.\ref{f1}, 
which mimics very tightly the $R$-boost solution. 
Indeed in this phase the Q kinetic energy density 
dominates and Eq.(\ref{phiRDEapprox}) implies 
\begin{equation}
\label{kinRDEapprox}
{1\over 2}\phi_{t}^{2}={1\over 2}{\dot{\phi}\over a^{2}}=
{1\over 2}{\phi_{beg}^{2}{\cal C}^{2}\over a^{2}}\ .
\end{equation}
The dot-dashed line in Fig.\ref{f1} has been obtained 
by inserting the value of ${\cal C}$ and $\phi_{beg}$ 
at the onset of the $R$-boost. 

Note that the scaling as $1/a^{2}$ is sensibly different
from a truly kinetic dominated phase in minimally coupled models, for
which we should have had $1/a^{6}$, from the continuity
equation (\ref{continuity}) with $p_{\phi}\simeq\rho_{\phi}$.
The reason why we have a different scaling of
$\rho_{\phi}$ is that the NMC terms in ${\tilde\rho}_{\phi}$
are not negligible during all the RDE and the first part of
the MDE, as we will show in a moment. Let us conclude the
description of the $R$-boost by estimating the time of its
end. The latter occurs at a redshift $z_{pot}$ when the true
potential energy becomes comparable with the kinetic energy
(\ref{kinRDEapprox}):
\begin{equation}
\label{Rboostend}
{1\over 2}\phi_{beg}^{2}{\cal C}^{2}(1+z_{pot})^{2}=
V[\phi(z_{pot})]\ .
\end{equation}
Since the $R$-boost is a very high redshift process, it
covers a very tiny time interval, and in practice $\phi$
does not move substantially from its initial condition
$\phi_{beg}$ during this phase. Therefore we can take
$V[\phi(z_{pot})]\simeq V(\phi_{beg})$ in Eq.(\ref{Rboostend}).
Moreover, for our inverse power-law potential we have
$V(\phi_{beg})=V(\phi_{0})(\phi_{0}/\phi_{beg})^{2}$, and
from the Friedmann  equation today $8\pi G V(\phi_{0})/3
=\Omega_{\phi}H_{0}^{2}$. Putting these ingredients together
we can write an approximate formula for $1+z_{pot}$
giving the end of the $R$-boost:
\begin{equation}
\label{Rboostendformula}
1+z_{pot}\simeq\sqrt{6\Omega_{\phi}H_{0}^{2}\phi_{0}^{2}
\over 8\pi G{\cal C}^{2}\phi_{beg}^{4}}\simeq 3000\ ,
\end{equation}
where the precise  number has been obtained by making use of
the actual value of the parameters in our case: it
correctly corresponds to the $R$-boost end in
Fig.\ref{f1}. Note that the end of the $R$-boost actually occurs when
the Universe has already become matter dominated.  

Before concluding this subsection, let us return to the
importance of the NMC terms in $\tilde{\rho}_{\phi}$ in
equation (\ref{rhophi}); as we already anticipated, we show
now that they are dominant with respect to kinetic and potential
energy densities until the first part of the MDE. Indeed, the NMC
term in $\tilde{\rho}_{\phi}$ is
$-3{\cal H}F_{\phi}\dot{\phi}/a^{2}$ and
scales roughly as $1/a^{3}$, since ${\cal H}$ goes
like $1/a$; on the contrary, the true potential
$V$ is roughly constant and the $R$-boost
kinetic energy scales like $1/a^{2}$; as we explained.
Because of these scalings, we expect that there exists
a time $\tau_{NMC}$ such that for $\tau <\tau_{NMC}$
the NMC term is larger than both the kinetic and potential energy,
while it becomes subdominant after this time. In fact,
assuming that the time dependence of the field is as in
the $R$-boost solution (\ref{phiRDEapprox}),
a simple calculation shows that the NMC energy density
term $-3{\cal H}F_{\phi}\dot{\phi}/a^{2}$
dominates $\rho_{\phi}$ up to 
\begin{equation}
\label{nmcdominationpot}
\tau_{NMC} \simeq\left[{3\xi{\cal C}\phi_{beg}^{2}\over
\Omega_{\phi}\rho_{r0}H_{0}^{3}}
\left({\phi_{beg}\over\phi}\right)^{2}\right]^{1/3} \ \
{\rm or}\ 1+z_{NMC} \simeq 850\ ,
\end{equation}
that is long after matter radiation equivalence, as
it is evident in Fig.\ref{f1}.
Note however that this is only an approximation,
since Eq.(\ref{Rboostendformula}) tells us that
the $R$-boost solution is no longer satisfied at these
redshifts; anyway this clearly shows that
the NMC terms become subdominant only after the onset
of the matter domination, so that the distinction between
$\rho_{\phi}$ and ${\tilde\rho}_{\phi}$ can be relaxed after
$\tau_{NMC}$. Our analysis is therefore approaching
the MDE behavior of $\phi$, subject of the next subsection.

\subsection{TEQ-trajectories in the Matter Dominated Era}
 
Tracking solutions have been recently obtained
in the context of Induced Gravity models \cite{Uzan}, whose only
difference compared with NMC is the absence of  
a constant term multiplying the Ricci scalar 
in the gravitational sector of the Lagrangian, i.e. 
$F(\phi )=\xi\phi^{2}$. We make two claims here. 
The first is that the same tracker solutions also exist in NMC models, 
provided in the MDE the scale factor can be expressed as a power of
$\tau$. The second is that for $F(\phi)\propto \phi^{\beta}$ the same 
solutions exist only if $\beta =2$, as assumed so far. 

In fact, by looking at Eqs.(\ref{Friedmann}, \ref{KG}), 
it is immediately realized that the difference between 
the IG and NMC models resides only in the $1/3F$ term 
multiplying the RHS in (\ref{Friedmann}), 
since all the other terms involving $F$ are derivatives 
with respect to either time or $\phi$. Tracking 
solutions in \cite{Uzan} have been obtained 
by assuming that the scale factor in the MDE
is the square of the conformal time $\tau$, 
\begin{equation}
\label{atrack}
a(\tau)=a_{*}\left({\tau\over\tau_{*}}\right)^{2}\ ,
\end{equation}
($\tau_{*}$ is a generic time) 
so that Eq.(\ref{Friedmann}) gets almost identically satisfied and disappears
from the treatment. Indeed we show that in NMC models both
Eq.(\ref{Friedmann}) and Eq.(\ref{atrack}) hold true, simply because 
matter is dominating and $F(\phi )$, playing the role 
of the gravitational constant in Eqs.(\ref{Friedmann},
\ref{Friedmann2}), is slowly moving. 
Since $\phi_{beg}\ll\phi_{0}\simeq 0.35M_{P}$, it is easy 
to see that the variation of the value of $F$ throughout 
the whole cosmological trajectory is 
\footnote{It is also useful to point out here that, because of the 
small change in the value of $F(\phi)$ throughout the entire
cosmological evolution, the well-known constraints on the 
variation of $H$ at the epoch of nucleosynthesis are always satisfied 
in our models (see e.g. \cite{PBM} and references therein).} 
\begin{equation}
\label{deltaF}
\left|{F_{0}-F_{beg}\over F_{0}}\right|
\simeq 4.6 \times 10^{-2}\ .
\end{equation}
To see this in another way, Fig.\ref{f3} shows the behavior 
of ${\cal H}$, indicated on the vertical axis as $(1+z)H^{-1}$, 
where $H=a_{t}/a$, in three important moments 
of the cosmological evolution, namely present (top), 
decoupling (middle), and equivalence (bottom); the heavy, solid 
line represents TEQ with positive $\xi$, the thin solid line is TEQ with 
negative $\xi$, the short dashed line, instead, is ordinary Q, and the
long dashed one is the cosmological constant, which is sensibly different from 
Q and TEQ since it is not dynamical. As it is evident, opposite 
signs of $\xi$ imply  opposite behaviors for ${\cal H}$, especially at  
small redshifts, compared with ordinary Q. 
At a given redshift, the shift in ${\cal H}$ is due to the 
behavior of $F(\phi )$, which is less or more than $1/8\pi G$ 
for positive or negative $\xi$, respectively, by an amount 
given by (\ref{deltaF}). 

Therefore, Eq.(\ref{atrack}) holds with good 
accuracy, and all the solutions obtained in 
\cite{Uzan} for IG models, as well as their stability 
properties, hold true also in our NMC case. 

Let us come now to our second claim. Let us assume for a 
while a form $F(\phi)=\tilde{\xi}\phi^{\beta}$ and $F(\phi )=
1/(8\pi G)+\tilde{\xi}(\phi^{\beta}-\phi_{0}^{\beta})$, 
respectively.  
We are searching for power law solutions for the Quintessence 
\begin{equation}
\label{phitrack}
\phi (\tau )=\phi_{*}\left({\tau\over\tau_{*}}\right)^{\gamma}\ ;
\end{equation}
from (\ref{atrack}) and (\ref{phitrack}), 
it is matter of simple algebra to check that 
Eq.(\ref{KG}) takes the form 
\begin{equation}
\label{KGtrack}
\left(\gamma^{2}+3 \gamma\right){\phi\over \tau^{2}}=
{a^{2}R\over 2}\beta\tilde{\xi}\phi^{\beta -1}+a^{2}\alpha
{M^{4+\alpha}\over\phi^{\alpha +1}}\ .
\end{equation}
Also, from (\ref{riccih}) the Ricci scalar in the MDE 
becomes $R=12/(a\cdot\tau)^{2}$. 
It is easy to see that this is only satisfied if 
\begin{equation}
\label{KGtracksol}
\beta =2\ \ {\rm and}\ \ \alpha ={6\over\gamma}-2\ . 
\end{equation}
This proves our second claim. 
If conditions (\ref{KGtracksol}) 
are satisfied, Eq.(\ref{KGtrack}) gives $\phi_{*}$ 
in terms of $\gamma,n,\alpha$; since in our models 
$\alpha >0$ and the field is positive, 
the following additional condition is derived: 
\begin{equation}
\label{additionaltrack}
\gamma^{2}+3 \gamma - 12\tilde{\xi} >0\ .
\end{equation}
In this regime, the Quintessence energy density scales as 
\begin{equation}
\label{rhotrack}
\rho_{\phi}=\rho_{\phi *}\left({a\over a_{*}}\right)^{-\epsilon}
\ \ ,\ \ \epsilon = {3\alpha\over \alpha + 2}\ ,
\end{equation}
and its pressure is 
\begin{equation}
\label{ptrack}
p_{\phi}=-\rho_{\phi}{2\over \alpha +2}\ .
\end{equation}
For what concerns the stability of these solutions, 
we report here only a remarkable result \cite{Uzan}; 
solutions satisfying (\ref{KGtracksol}) are stable 
under time dependent perturbations if and 
only if the following condition holds: 
\begin{equation}
\label{stabilitytrack}
-1-{4\over\alpha +2}<0\ ,
\end{equation}
holding for instance for any $\alpha >0$. 
In summary, for the class of solutions we are interested in, 
namely those satisfying (\ref{atrack},\ref{KGtrack}), the cases 
of interest here have been treated in \cite{Uzan}; we have shown 
here that such solutions apply both to NMC and IG cases. 

Let us describe now in detail the solutions we see 
in Fig.\ref{f1}. 
Let us focus on the minimal coupling case first. 
The lowest, dashed line 
represents a minimally coupled case in which the behavior of 
the Q component during RDE is equivalent to a pure 
cosmological constant. 
The kinetic energy density is initially zero and remains largely 
subdominant with respect to the potential one. 
Thus $\phi$ ``freezes" for almost all the RDE, and leaves this 
condition only when its energy density reaches a fraction of 
about $10^{-3}$ of the critical one. Thereafter, Q joins 
the tracker solution in the MDE regime corresponding to 
\begin{equation}
\gamma = \epsilon = 1.5\ ,
\label{exptrack}
\end{equation}
that satisfies the constraint (\ref{additionaltrack}) with 
$\tilde{\xi}=0$. The substantial motion of the field from 
the initial condition to its final value occurs in this last 
phase. 

The thin long-dashed line represents a case in which the initial 
kinetic energy density is dominant with respect to the 
potential one, by 23 orders of magnitude. As it is 
evident, the field starts from an energy density comparable 
with the matter one. In this case 
the kinetic energy is redshifted away during the 
phase corresponding to the rapidly decreasing part of the trajectory 
in the figure; the scaling is easily found from the continuity equation,  
with $\rho_{\phi}\simeq p_{\phi}$. 
Then the field freezes again before joining the tracker 
solution at $1+z\simeq 100$. Note that because of this 
early stage of kinetic energy dominance, the field 
freezes at a value slightly greater than the initial condition 
$\phi_{beg}=10^{-2}$; that is the reason why the flat 
part of the curve lies slightly below that corresponding to
the previous case. Although not plotted in the 
figure, a roughly equivalent trajectory might have been 
obtained by requiring that the initial potential energy 
density was comparable to the matter one. 
Note the very large set of initial 
energy values from which the field reaches the present state. 

Let us come now to the analysis of the tracker solutions 
in the NMC case. The heavy solid line has been obtained for 
$\xi =1.5\times 10^{-2}$. The RDE is dominated by the $R$-boost: 
the field accelerates until the gravitational effective 
potential is reached by the cosmological friction term 
in the Klein-Gordon equation. The slow rolling sets in, 
and holds until the true potential becomes important, when 
the field freezes. After this early phase, 
${\tilde\rho}_{\phi}$ and $\rho_{\phi}$, after having been much 
different in magnitude and sign, become indistinguishable, 
and join the tracker solution in the MDE 
reaching the required value today.  

The heavy dashed line corresponds to a case in which 
the initial kinetic energy density is larger by 23 
orders of magnitude with respect to the potential one, 
thus starting from an energy amount comparable with 
the matter one. As for the Q case, in this condition 
the field undergoes an era of kinetic energy dominance 
until the latter is redshifted below the effective 
NMC potential energy and the $R$-boost is set also in this 
case. The evolution of the field from this time on is the 
same as in the zero initial kinetic energy case. 

The dot-dashed line is in fact the $R$-boost 
approximate solution (\ref{phiRDE}). Note that it describes 
very well the $R$-boost, and shows as the terms of order higher 
than the first in Eq.(\ref{phiRDEapprox}) become important only 
now, when the $R$-boost is no longer active. 

The last point we want to stress, 
is that the heavy solid line is actually superposed with a solid 
thin line, which describes a case identical, 
but for negative coupling constant, $\xi =-1.5 \times  10^{-2}$. 
The reason why these two trajectories agree so tightly is that during
the $R$-boost 
the kinetic energy density, which dominates $\rho_{\phi}$, is 
the same regardless of the sign of $\xi$, see Eqs. (\ref{phiRDE}-
\ref{kinRDEapprox}). The only slight difference between the two 
trajectories is when the $R$-boost ends and the true potential 
starts to drive the dynamics of $\phi$. For positive $\xi$, 
$\dot{\phi}$ is positive during the $R$-boost and later, while, 
for negative $\xi$, $\dot{\phi}$ is negative during the $R$-boost.
Although the curves describing $\rho_{\phi}$ for positive and 
negative values of $\xi$ look very similar, this is not true 
for the perturbations they generate, as will be discussed  
in the next section. 

\section{Cosmic Microwave Background spectra}

In this section we analyse the CMB temperature and 
polarization spectra of our TEQ model.  

In Fig.\ref{f4} we plot the CMB spectra 
for the models discussed in the previous 
section; the top panel describes temperature fluctuations; the
bottom panel shows polarization spectra. 
The CMB spectrum coefficients are 
calculated from
\begin{equation}
\label{cl}
C_{\ell}^{T}=
4\pi\int{dk\over k}|\Delta_{T\ell}(k,\tau_{0})|^{2}
\ \ ,
\ \ C_{\ell}^{P}=
4\pi\int{dk\over k}|\Delta_{P\ell}(k,\tau_{0})|^{2}\ ,
\end{equation}
where the quantities $\Delta_{T\ell}(k,\tau_{0})$ and
$\Delta_{P\ell}(k,\tau_{0})$ are functions of the 
photon and baryon perturbed quantities, see e.g.  
ref. \cite{CF} for detailed definitions. 
The spectra have been normalized at the COBE measurements at $\ell =10$. 

The heavy, solid line in Fig.\ref{f4} describes CMB spectra for the
model corresponding to the same line in Fig.\ref{f1}. The thin, solid line 
describes the same model, but for $\xi =-1.5\times 10^{-2}$, 
again as the same line in Fig.\ref{f1}. The
short-dashed line, in the left panels, represents 
a case of ordinary Quintessence with the same potential and 
$\Omega_{\phi}=0.7$. The long-dashed line, in the right 
panels, describes a cosmological constant, $\Lambda$ model, 
with $\Omega_{\Lambda}=0.7$. Before entering into the description 
of the various effects we find, let us stress that 
the effects of TEQ are quite large, with respect to both ordinary Q 
and $\Lambda$ models. Also, the models plotted respect  
the experimental constraints discussed in Section II, having 
$\omega_{JBD}\simeq 500$ and $G_{t}/G\simeq 10^{-12}$ yr$^{-1}$. 
We want to mention that the effects are here considerably larger 
than what we obtained in our previous paper \cite{PBM}; the reason is
twofold, in \cite{PBM} we did not follow the tracker solution for the 
Q field, but for the shallower potential we considered there ($\alpha=1$) 
the size of the effects we find here would have been smaller anyway 
(see the discussion in Section VI). 

\begin{figure} 
\centerline{
\psfig{file=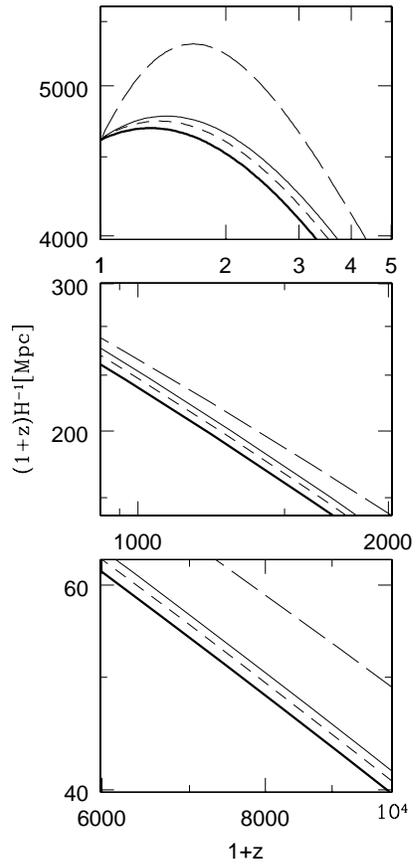,height=5.in,width=5.in}
}
\caption{Snapshots of conformal Hubble horizon 
evolution. heavy solid: Tracking Extended Quintessence 
with positive $\xi$; 
short dashed: tracking minimally coupled Quintessence;  
thin solid: Tracking Extended Quintessence  
with negative $\xi$; long dashed: cosmological constant. 
Top: present; middle: decoupling; bottom: equivalence.}
\label{f3}
\end{figure}

\begin{figure} 
\centerline{
\psfig{file=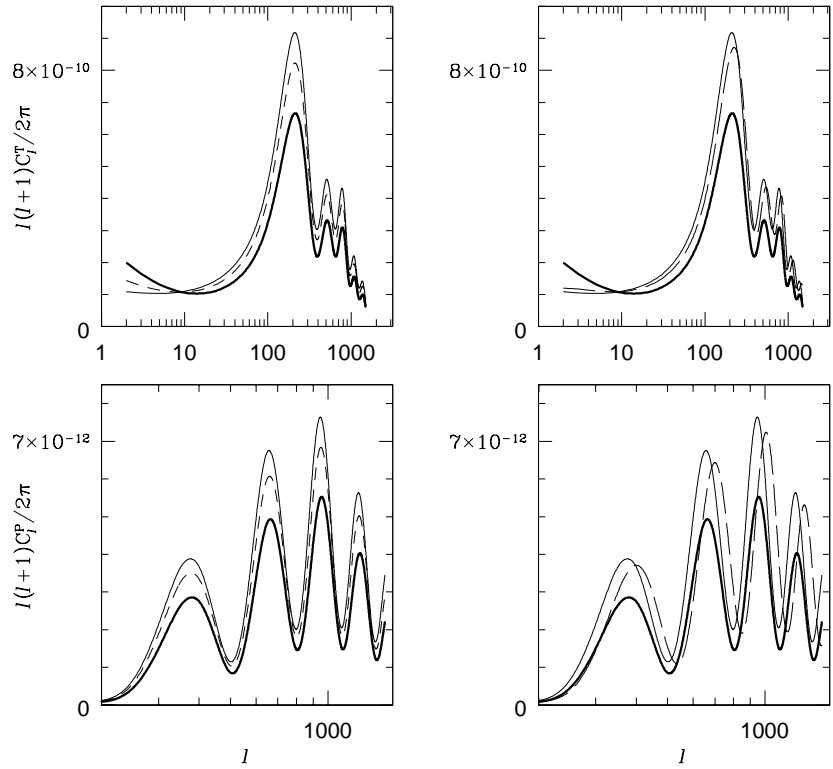,height=5.in,width=5.in}
}
\caption{CMB angular power spectra. Top: temperature; bottom
polarization. 
Heavy solid: Tracking Extended Quintessence, positive $\xi$; 
thin solid: Tracking Extended Quintessence, negative $\xi$; 
short dashed: minimally coupled tracking 
Quintessence; long dashed: cosmological constant.}
\label{f4}
\end{figure}

First of all, let us consider the dynamics of Q at low 
redshifts. 
By looking at Eq.(\ref{KG}), it is easily seen that 
at present Q obeys a sort of effective potential,  
caused by the true potential and by the curvature-coupling term: 
\begin{equation}
\label{poteff}
V_{eff}(\phi ,R)=V(\phi )-{F(\phi )R\over 2}\ .
\end{equation}
At low redshifts when Q starts to dominate, the 
dynamics of the Hubble length is suppressed since 
the universe is approaching a de Sitter phase. 
Therefore, by neglecting $\dot{\cal H}$ in 
Eq.(\ref{riccih}), we have $R\simeq 6H^{2}$. 
Also, from the Friedmann equation we get 
$(8\pi G/3)V\simeq \Omega_{\phi}H^{2}$, and 
for inverse power-law potentials of the form 
(\ref{v}) we have $dV/d\phi=-\alpha V/\phi$. 
Thus the derivative of the effective potential 
(\ref{poteff}) takes the following approximate form: 
\begin{equation}
\label{dpoteffdphi}
{dV_{eff}\over d\phi}\simeq - 
\left({3\alpha\Omega_{\phi}\over 8\pi G\phi}+
6\xi\phi\right)H^{2}\ ;
\end{equation}
for positive $\xi$, both terms push toward increasing 
values of $\phi$, and we can immediately understand that 
the NMC term is comparable to the ordinary Quintessence 
one for 
\begin{equation}
\label{NMCISW}
|\xi |={\alpha\Omega_{\phi}\over 16\pi G\phi^{2}}\ .
\end{equation}
Therefore, for $\xi =1.5\times 10^{-2}$ and $\alpha =2$, 
we expect to have a $10\%$ extra force 
coming from the NMC terms in Eq.(\ref{Friedmann}). 
Just the opposite happens if $\xi$ is negative, since 
the terms in (\ref{dpoteffdphi}) have opposite sign. 
Indeed what we see in Fig.\ref{f3} is that 
for opposite signs of $\xi$, the change in the low redshift 
dynamics of ${\cal H}^{-1}$ is also opposite. 
Also, in Fig.\ref{f2} it can be seen that at present 
roughly an order of magnitude separates the gravitational 
effective potential from the true one. 

The Integrated Sachs Wolfe effect (ISW, see e.g. \cite{HSWZ} 
for a review) makes the CMB coefficients on large scales, 
or small $\ell$'s, change with the variation of the gravitational 
potential along the CMB photon trajectories. 
Since the gravitational 
potential is affected by the low redshift dynamics, we expect 
an increase or a decrease of the ISW for positive and 
negative $\xi$, respectively. Indeed this is precisely what 
we see in Fig.\ref{f4}, looking at the curves for 
$\ell\lesssim 10$: positive or negative $\xi$ TEQ make the $C_{\ell}$ 
larger or smaller than ordinary Q, respectively. 

The following simple calculation gives a good estimate of 
the amount of ISW in TEQ models. 
Take a cosmological scale comparable with the Hubble 
horizon today, so to be unaffected by acoustic 
oscillations. As it is known 
(see e.g. \cite{HSWZ} and references therein), 
the ISW is essentially due to the change, between decoupling and 
now, of the gauge-invariant expression of the gravitational potential, 
$\Psi$. In NMC theories $\Psi_{dec}$ 
is slightly different from the one calculated in  
minimally coupled theories, because it receives a 
contribution from the time variation of the 
gravitational constant, since $\Psi$ is proportional 
to $G \propto 1/F$; going from decoupling to now, 
an increasing $F$ (positive $\xi$) makes $\Psi$ 
decreasing, while the opposite is true for 
decreasing $F$. More precisely, 
in the limit $|F_{dec}-F_{0}|\ll F_{0}$, $\Psi$ changes 
by an amount 
\begin{equation}
\label{deltapsi}
\delta\Psi =\Psi \left({F_{0}\over F_{dec}}-1\right)\ .
\end{equation} 
The more the gravitational potential decreases, the 
stronger is the ISW power at $\ell \le 10$, which is 
roughly given by $(\delta T/T)_{ISW}\simeq 2\delta\Psi$ 
\cite{HSWZ}. 
Therefore, recalling that on superhorizon scales 
$\delta T/T\simeq \Psi /3$, on the CMB power spectrum 
the NMC contribution to the ISW can be 
estimated as 
\begin{equation}
\label{iswnmccl}
{\delta C_{\ell\lesssim 10}\over C_{\ell\lesssim 10}} = 
{C_{\ell\lesssim 10}^{TEQ}-C_{\ell\lesssim 10}^{Q}\over 
C_{\ell\lesssim 10}^{Q}}= 
{(\delta T/T)_{ISW,NMC}^{2}-(\delta T/T)_{ISW}^{2}
\over (\delta T/T)_{ISW}^{2}}\simeq 
12\cdot{F_{0}-F_{dec}\over F_{0}}
\simeq 96\pi G\xi\phi_{0}^{2}\ ;
\end{equation}
the last equality has been obtained in our 
particular model, where at decoupling $\phi\ll\phi_{0}$, 
so that $1-F_{dec}/F_{0}\simeq 8\pi G\xi\phi_{0}^{2}$. 
Note again that, depending on the sign of $\xi$, 
the net effect can be an increase or a decrease of the CMB 
power. Indeed in Fig.\ref{f3} 
we see that the above estimate is fairly well respected: 
since in the present case $\phi_{0}\simeq 0.35M_{P}$ on large 
angular scales Eq.(\ref{iswnmccl}) predicts 
an effect of the order of $10\%$, as in the figure. 

Let us come now to the evaluation of the effects on the 
acoustic peaks in the CMB spectrum. There are essentially 
two effects, peaks amplitude and position, that we describe now. 

The amplitude changes in the opposite way for opposite sign  
of $\xi$. This feature can be understood as a normalization 
effect: we have seen that positive and negative $\xi$ give rise 
to increased and decreased ISW in TEQ with respect to Q models; 
correspondingly, when the spectra are normalized at $\ell =10$, 
a decrease and increase of the acoustic peak amplitude occurs. 
As a consequence, the magnitude of this effect is roughly at the 
level given by Eq.(\ref{iswnmccl}). 
For completeness, 
we report also another mechanism that changes 
the acoustic peaks amplitude, which occurs at decoupling,  
instead of at low redshifts. As is evident in Fig.\ref{f3}, 
the size of the Hubble horizon at a given redshift is different 
in TEQ with respect to Q models. As we noted in \cite{PBM}, 
this implies that the horizon reentry for a given 
comoving scale is delayed for positive $\xi$ and 
anticipated for negative $\xi$; therefore, with 
respect to the ordinary tracking Quintessence, this 
implies a slight excess or deficit in CDM density 
when such scale is in horizon crossing, corresponding 
to a deficit and an excess in the radiation energy density. 
This slightly suppresses the acoustic oscillations 
in the first case, and enhances them in the second. 

Let us come now to the explanation of the acoustic peaks shift. 
The angular scales at which acoustic oscillations occur are 
directly proportional to the size of the CMB sound horizon at 
decoupling, that in comoving coordinates is roughly 
$\tau_{dec}/\sqrt{3}$, and inversely proportional to the 
comoving distance covered by CMB photons from last scattering 
until observation, that is $\tau_{0}-\tau_{dec}$ \cite{HSWZ}. 
Therefore, since the spectrum multipoles scale as the inverse 
of the corresponding angular scale, we have 
\begin{equation}
\label{lpeaks}
\ell_{peaks}\propto {\tau_{0}-\tau_{dec}\over\tau_{dec}}\ .
\end{equation}
In our models, $\tau_{0}$ and $\tau_{dec}$ shift essentially 
because of two reasons, namely the change in $\tau_{0}$ due 
to the domination of the Q component today, and the behavior 
the Hubble length $H^{-1}$ in the past. The first feature 
mainly makes the Q and TEQ models different from a cosmological 
constant, since the latter does not have the kinetic degree of freedom. 
The second feature is simply related to the fact that, for a fixed 
value of the Hubble length today, in the past it possesses 
a behavior which is characteristic of the particular model at hand, 
because of the time evolution of the effective gravitational 
constant $1/F(\phi )$ (see Fig.3). 
Let us give a simple analytical formula for 
the acoustic peaks shift in TEQ models with respect to ordinary 
tracking Q. First, it can easily be seen that 
the conformal time $\tau$ can be conveniently written 
as a function of the scale factor as follows: 
\begin{equation}
\label{taua}
\tau =\int_{0}^{a}{da\over a\dot{a}}\ .
\end{equation}
However, from the Friedmann equation (\ref{Friedmann}), 
$\dot{a}$ scales as the inverse of $\sqrt{F}$; 
therefore, small changes $\delta F\ll F$ induce a change 
$\delta\dot{a}/\dot{a}\simeq -(1/2)\delta F/F$, and 
consequently in the conformal time, which shifts by an 
amount 
\begin{equation}
\label{taushift}
\delta\tau=\int_{0}^{a}{da\over a\dot{a}[1-(1/2)\delta F/F]}-\tau
\simeq {1\over 2}\int_{0}^{a}{da\over a\dot{a}}{\delta F\over F}\ ,
\end{equation}
where we have defined the time change of $F$ as follows: 
\begin{equation}
\label{dff}
{\delta F\over F}={F(\phi )-F_{0}\over F_{0}}\ .
\end{equation}
Of course $\delta F/F$ at any given time depends 
on various details, but let us make the 
simplifying assumption that it is constant from 
the beginning until $z=2$ and zero afterwards, 
empirically following what we see in 
Fig.3. It is then immediate to deduce that 
$\tau_{dec}$ changes as 
\begin{equation}
\label{dtaudec}
\delta\tau_{dec}\simeq 
\tau_{dec}\cdot {1\over 2}{\delta F\over F}\ .
\end{equation}
Instead, for what concerns $\tau_{0}$ we have 
\begin{equation}
\label{dtau0}
\delta\tau_{0}\simeq 
\tau_{z=2}\cdot {1\over 2}{\delta F\over F}\ ,
\end{equation}
because, according to our simplifying assumption, 
$\delta F/F \approx 0$ for $z \lesssim 2$. 
In conclusion, minding Eq.(\ref{lpeaks}), and after 
some algebra, we get the shift in the acoustic 
peaks as a result of the time variation of the 
effective gravitational constant: 
\begin{equation}
\label{dll}
{\delta l_{peaks}\over l_{peaks}}=
{l^{TEQ}_{peaks}-l^{Q}_{peaks}\over l^{Q}_{peaks}}\simeq 
{1\over 2}{\delta F\over F}
\left({\tau_{z=2}\over\tau_{0}}-1\right)\ .
\end{equation}
Numerically we find $\tau_{z=2}/\tau_{0}\simeq 75\%$; 
also, we already mentioned that the change of the value 
of $F$ during all the cosmological evolution can be written 
as $\delta F/ F\simeq -8\pi \xi G\phi_{0}^{2}$. 
Therefore, for our specific model Eq.(\ref{dll}) becomes 
\begin{equation}
\label{dllour}
{\delta l_{peaks}\over l_{peaks}}\simeq \pi\xi G\phi_{0}^{2}\ .
\end{equation}
Note that, for our values of $\xi =\pm 1.5\times 10^{-2}$, the 
above shift is at the level of $\pm 6\times 10^{-3}$, which 
is in quite good agreement with the results plotted 
in the left panels of Fig.4, that is $\pm5\times 10^{-3}$. 

This completes our description of the TEQ features on the 
CMB angular power spectrum. 
We turn now to the analysis of what happens in 
the matter power spectrum today. 

\section{Matter power-spectrum}

\begin{figure} 
\centerline{
\psfig{file=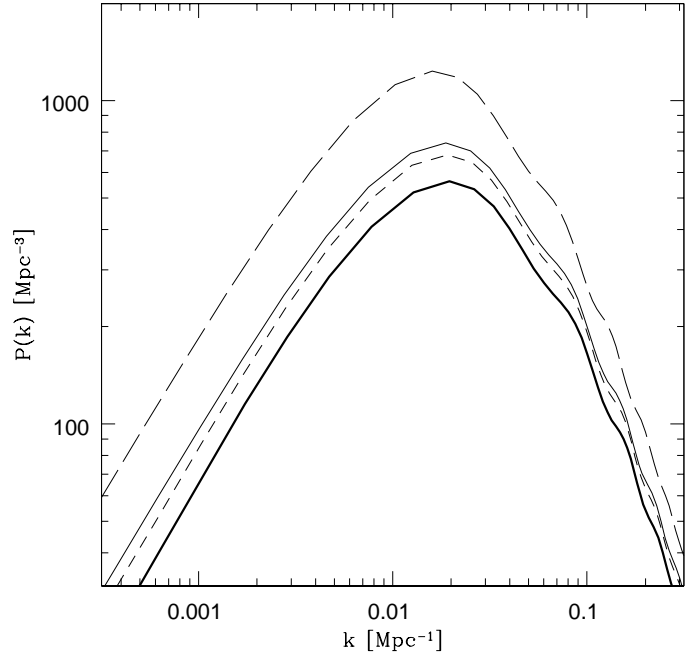,height=5.in,width=5.in}
}
\caption{Present matter power spectra. 
Heavy solid: Tracking Extended Quintessence, positive $\xi$; 
thin solid: Tracking Extended Quintessence, negative $\xi$; 
short dashed: minimally coupled tracking 
Quintessence; long dashed: cosmological constant.}
\label{f5}
\end{figure}

In Fig.\ref{f5} we plot the matter power spectrum for the same cases shown 
in Fig.\ref{f4}. We can note immediately differences regarding both  
amplitude and turnover position. 

The $\Lambda$ model has the highest spectrum. The main reason 
is the different growth of density perturbations 
\cite{CDF}. In both Q and $\Lambda$ models the perturbation 
growth is suppressed at low redshifts due to the 
domination of the vacuum energy, that tends to keep 
$H$ constant therefore 
enhancing the cosmological friction in the perturbation 
equations, where almost everywhere the terms involving 
the first time derivatives of the perturbations appear 
multiplied by the Hubble parameter. 
In Q and TEQ models this effect is considerably enhanced due 
to the magnitude of $H$ which is greater than in 
$\Lambda$ models at all redshifts, and in particular 
at the lowest ones, as it is evident from Fig.\ref{f3}. 
Another independent cause that contributes 
to push the Q spectrum down with respect to the 
$\Lambda$ model is the COBE normalization. In fact, 
we have seen that the ISW effect is enhanced in Q models 
with respect to $\Lambda$ ones, and the normalization 
to COBE implies subtraction of power to the true amplitude 
of the primordial cosmological perturbations. 

Let us come now to the difference between Q and TEQ models. 
The dynamics of the field at low redshifts is almost the 
same as in Q models, as it is evident again by looking 
at Fig.\ref{f1}. 
Thus, the reason of the difference is to be searched for
in the COBE normalization. Indeed, by looking at the 
low wavenumber region, which is the zone of 
non-processed scales, we see that the amplitude shift 
is roughly at $10\%$ and $-18\%$ level for TEQ with respect to 
Q for negative and positive $\xi$, respectively; these numbers 
roughly agree with the ISW corrections that we estimated 
to come from the NMC terms in the previous section. 

Let us come now to the evaluation of the peaks shifts. 
As it is known (see for example \cite{CDF}), 
the scale of the matter power spectrum 
turnover is essentially given by the scale entering 
the Hubble horizon at the matter-radiation equivalence. 
The latter age is the same for all our models: 
\begin{equation}
\label{eq}
1+z_{eq}={\rho_{m0}\over\rho_{r0}}\approx 5500 \ . 
\end{equation}
However, we must take care of what was the Hubble horizon at the
equivalence, since the Hubble radius follows different dynamics in the
three cases that we are treating. In other words, the shift in the power 
spectrum turnover is given by 
\begin{equation}
\label{peakshiftH}
{\delta k_{turn}\over k_{turn}}=-
\left({\delta H^{-1}\over H^{-1}}\right)_{eq}\ .
\end{equation}
In Fig.\ref{f3} the different values of the Hubble horizon 
are displayed at the equivalence (bottom). 
The Hubble horizon shift in Eq.(\ref{peakshiftH}) between the Q 
and $\Lambda$ model is at the $18\% $ level, which corresponds 
well to the power spectrum shift that we see in  
Fig.\ref{f5}, where $\delta k_{turn}/k_{turn}\simeq -20\%$. 
The same reasoning applies to explain the slight shift 
of the TEQ spectra turnover with 
respect to the Q one. The Hubble horizon shift in Fig.\ref{f3} 
for positive $\xi$ is $2.6\%$, in good agreement with what we get 
in Fig.\ref{f5}, $2.3\%$; for negative $\xi$ we get 
$1.7\%$ from both figures. 

We can give a rough analytical estimate of this 
effect by reasoning as follows. 
At the equivalence the Q energy density, 
both in Q and TEQ models, is negligible with respect to 
matter and radiation as it is evident in Fig.\ref{f1}. 
Eq.(\ref{Friedmann}) takes the form 
\begin{equation}
\label{Friedmannapprox}
{\cal H}^{2}\simeq {\rho_{fluid}\over 3F(\phi )}\ .
\end{equation}
Since in our model $F(\phi )$ is smaller or larger than $F(\phi_{0})$ 
if $\xi$ is positive or negative, respectively, this implies 
that the Hubble length $H^{-1}$ at the equivalence 
was different in TEQ models with respect to ordinary Q; 
the amount and the sign of the difference can be 
estimated roughly as follows: 
from Eq.(\ref{Friedmannapprox}) we have 
\begin{equation}
\label{Hubbleshift}
\left({\delta H^{2}\over H^{2}}\right)_{eq}\simeq 
\left({2\delta H\over H}\right)_{eq}=1-8\pi G\cdot 
F_{eq}(\phi_{eq})\ .
\end{equation}
Therefore, the matter power spectrum turnover shifts 
due to the NMC terms by the following amount: 
\begin{equation}
\label{peakshift}
\left(\delta k_{turn}\over k_{turn}\right)_{NMC}=-
\left({\delta H^{-1}\over H^{-1}}\right)_{eq}=
{1-8\pi G F_{eq}(\phi )\over 2}\simeq 4\pi G\xi\phi_{0}^{2}\ ,
\end{equation}
simply by taking $\phi_{eq}\ll\phi_{0}$ as it is today. 
In our models, the quantity in Eq.(\ref{peakshift}) is 
roughly $2\%$, that agrees quite well with the numbers 
we obtain. 

For what concerns the shape of the power spectrum, 
there is no significant differences between 
Q and TEQ models, since this could arise 
only when the scalar field becomes important 
at low redshifts, where however the dynamics 
in the two scenarios is very similar. 
Notice that the oscillations that can be seen on the decreasing branch at high 
wavenumbers in figure \ref{f5} are just a residual of the acoustic oscillations 
of the photon-baryon fluid at decoupling. 

\section{Variations in the potential slope}

Varying the power $\alpha$ in the potential (\ref{v})
implies varying the dynamics of $\rho_{\phi}$ according to 
Eq.(\ref{rhotrack}). For increasing slopes of the potential, 
we expect that 
the low redshift effects are enhanced correspondingly. 

\begin{figure} 
\centerline{
\psfig{file=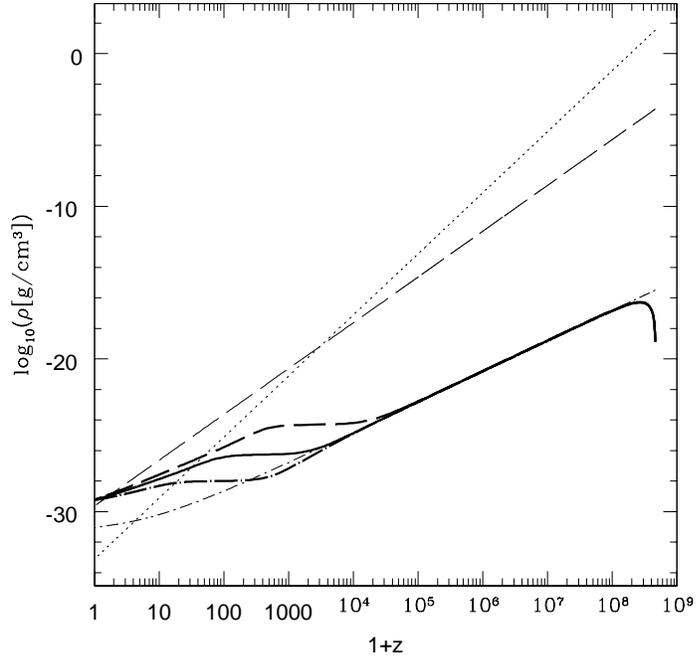,height=5.in,width=5.in}
}
\caption{Matter, radiation and Quintessence energy 
densities for different potential slopes. 
Long dashed: matter; short dashed: radiation; 
heavy solid: Tracking Extended Quintessence, $\alpha =2$;
heavy dotted dashed: Tracking Extended Quintessence, $\alpha =1$;
heavy long dashed: Tracking Extended Quintessence, $\alpha =3$.}
\label{f6}
\end{figure}

In Fig.\ref{f6} we show 
tracking solutions for different exponents: dot-dashed line
for $\alpha =1$, solid for $\alpha =2$, long-dashed for 
$\alpha =3$. It can be immediately seen how the low redshift 
tracker branch possesses different slopes, 
according to Eq.(\ref{rhotrack}). 
Moreover, the $R$-boost is the same for all the 
slopes, again according to the arguments made in 
Section II. But, for increasing $\alpha$, 
the $R$-boost is abandoned earlier. The reason is the mechanism that 
stops the $R$-boost itself, i.e. the fact that the Q potential energy 
starts to dominate. 
But, this happens earlier for larger exponents in Eq.(\ref{v}), 
since the potential is steeper for $\phi <\phi_{0}$. 

Another important aspect that we must address is the 
Q equation of state, $w_{\phi}=p_{\phi}/\rho_{\phi}$, 
since this is a parameter that is quite well constrained by the 
observations to be less than $-0.6$ today 
\cite{Perlm,Riess,Garna,Efsta}. 
In fact, from Eq.(\ref{ptrack}) 
we see that small $\alpha$ make the equation of state similar 
to the pure cosmological constant case. The actual value 
that we find numerically is slightly different from what 
Eq.(\ref{ptrack}) would predict, because today the tracker 
regime has been abandoned, since Q has started to dominate 
the cosmic evolution. For $\alpha = 1,2,3$, 
we find $w_{\phi}=-0.75$, $-0.60$ and $-0.49$, respectively, for
positive $\xi$, and $w_{\phi}=-0.78$, $-0.66$ $-0.59$, respectively, 
for negative $\xi$. 

\begin{figure} 
\centerline{
\psfig{file=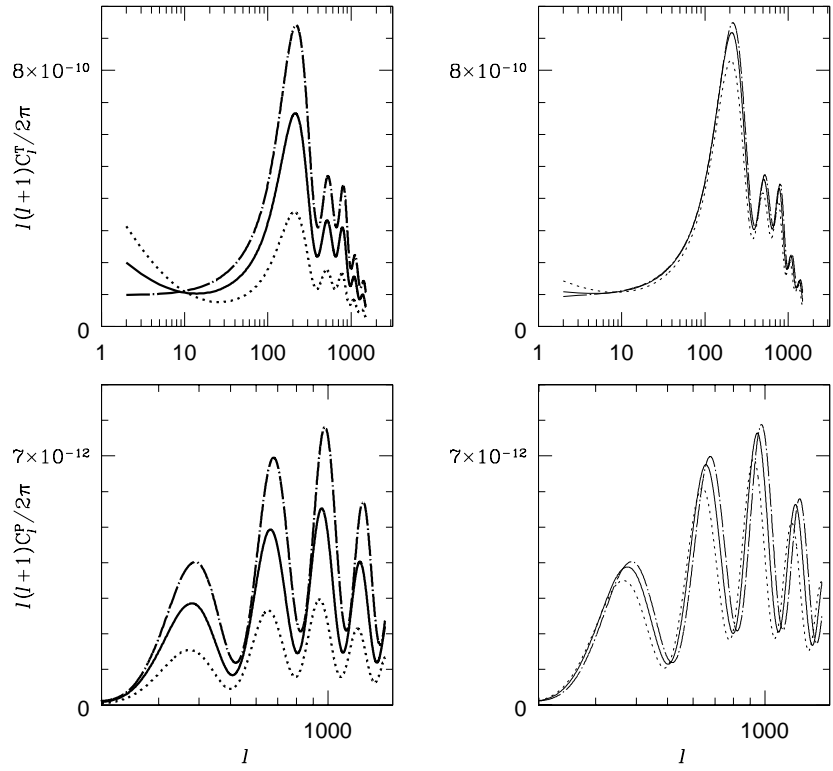,height=5.in,width=5.in}
}
\caption{CMB angular power spectra for different 
potential slopes. Left (right): positive (negative) $\xi$. 
Solid: Tracking Extended Quintessence, $\alpha =2$;
dotted dashed: Tracking Extended Quintessence, $\alpha =1$;
short dashed: Tracking Extended Quintessence, $\alpha =3$.}
\label{f7}
\end{figure}

Next, let us see what happens in the CMB. Fig.\ref{f7} shows 
the behavior of the CMB temperature (top) and polarization (bottom) 
spectra for the same values of $\alpha$ as in Fig.\ref{f6}; 
the left and right panels are for positive and negative $\xi$, 
respectively. 
Solid lines, both heavy and thin, refer to the same cases as in 
Fig.\ref{f4}; the dotted and dot-dashed lines describe 
models with $\alpha=3$ and $1$, respectively. 
As we mentioned above, the low redshift ISW effect is 
enhanced by increasing $\alpha$, merely because the 
low redshift dynamics is enhanced, as it is evident 
from Eq.(\ref{dpoteffdphi}). This is a known effect 
since it does not involve directly NMC terms, but 
instead is a characteristic feature of dynamical 
vacuum energy models \cite{Stain1,Stain2}. The acoustic peaks 
region is suppressed correspondingly, by an amount 
that is roughly increasing linearly with $\alpha$, 
due to the COBE normalization at low $\ell$'s. 
However, as it can be easily noted in Fig.\ref{f7}, 
the strength of this effect is larger for positive 
$\xi$ compared with negative values. The reason is 
that negative $\xi$ values tend to decrease the 
ISW effect, as it is evident both in the effective 
potential (\ref{dpoteffdphi}) and in our estimate of
Eq.(\ref{iswnmccl}). 
For what concerns the shift in the position of 
the acoustic peaks, the different potential mainly 
influences the low redshift tracking behavior, 
thus affecting mostly $\tau_{0}$ in (\ref{lpeaks}); 
since large $\alpha$s enhance the low redshift 
dynamics, the peaks shift increases correspondingly, 
again as it is evident in Fig.\ref{f7}. 

\begin{figure} 
\centerline{
\psfig{file=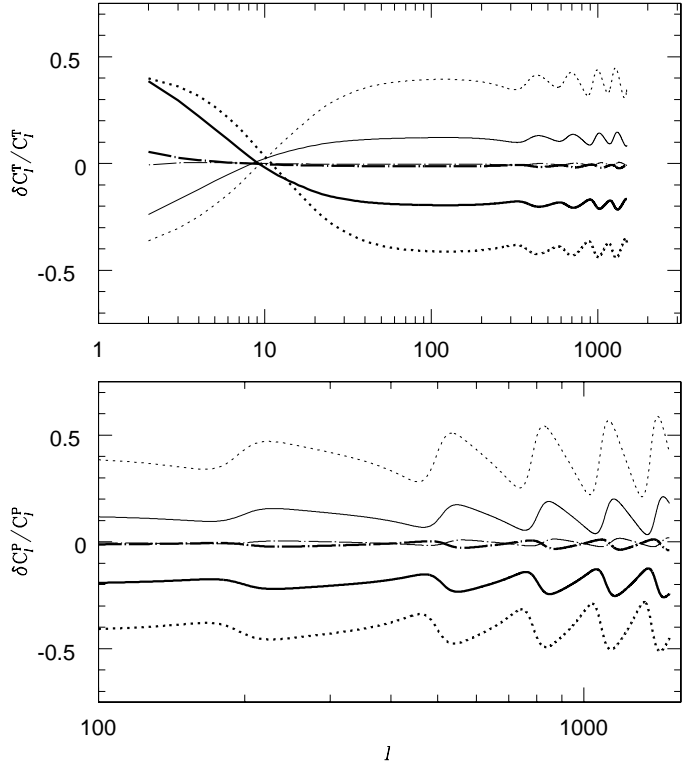,height=5.in,width=5.in}
}
\caption{CMB angular power spectra: relative difference between
Tracking Extended Quintessence and ordinary tracking Quintessence spectra
($\delta C_{\ell}/C_{\ell}=C_{\ell}^{TEQ}/C_{\ell}^{Q}-1$)
as a function of the inverse power of $\phi$ in the potential. Top (bottom):
temperature (polarization); thick (thin) lines for positive
(negative) $\xi$. Solid: extended vs. ordinary
tracking Quintessence, $\alpha =2$; dotted dashed for $\alpha =1$,
short dashed for $\alpha =3$.}
\label{f8}
\end{figure}

In conclusion, the increased 
dynamics obtained by increasing $\alpha$ has the indirect effect 
of enhancing all the NMC effects. To see this, 
let us stress the part of the effects that come 
from genuine NMC terms. To this aim we build tracking 
solutions corresponding to those that we show in 
Fig.\ref{f6}, but for the ordinary Q case, and calculate the CMB 
spectra for them. We take the ratios  
Eq.(\ref{iswnmccl}), 
now valid for all the $\ell$'s, 
\begin{equation}
\label{dclcl}
{\delta C_{\ell}\over C_{\ell}}={C_{\ell}^{TEQ}-C_{\ell}^{Q}\over
C_{\ell}^{Q}}\ ,
\end{equation}
to quantify in detail the pure NMC effects in the spectra. 
These quantities, for the different values of $\alpha$ treated 
in this Section, are displayed as a function of the multipole 
$\ell$ in Fig.\ref{f8}, temperature (top) and polarization (bottom). 
The meaning of the curves is the same as in Fig.\ref{f7}. 
The effects are larger for increasing values of $\alpha$. 
In the upper box we can note the large ISW effect due 
to the NMC component only: it reaches roughly $30\%$ 
by itself. Correspondingly, due to the normalization 
at small $\ell$'s, the acoustic peaks region is 
suppressed or enhanced for positive or negative $\xi$, respectively,  
even up to the $50\%$ level. 
The enhancement of these effects, in particular the ISW one, 
for increasing $\alpha$ in in agreement with our 
formula (\ref{iswnmccl}), since for increasing 
$\alpha$ the field reaches larger values today: 
namely, for $\alpha =1,2,3$, we have $\phi_{0}=0.2,0.35$ 
and $0.5$ Planck masses for positive $\xi$, and similar 
values for negative $\xi$. 

The same phenomenology obtained for the temperature anisotropies
is found for the polarization (bottom panels); we will not plot 
the low $\ell$'s region, since even if the NMC effect is large, 
the absolute value of the polarization multipole moments 
is significantly high only around degree angular scales. 

\section{Conclusions} 

In this section we will make a self consistent 
summary of the results obtained in this paper, focusing 
on the observable features of the cosmological models 
here investigated. 

The present paper is devoted to the study of 
theoretical aspects and observational imprints of cosmological scenarios 
in which a dynamical vacuum energy component dominates the 
cosmic evolution today, as suggested by high-redshift type Ia
Supernovae data \cite{Perlm}. 

Extended Quintessence means the class of models for which 
the vacuum energy today is given by a scalar field 
which has an explicit coupling with the Ricci scalar $R$; we proposed 
this scenario in \cite{PBM} and performed a preliminary study of its
cosmological properties. 
In the present work we assume a Non-Minimal Coupling (NMC) term
of the form 
\begin{equation}
\label{summaryf}
{R\over 16\pi G}+
\xi\left(\phi^{2}-\phi_{0}^{2}\right)\ ,
\end{equation}
where $\xi$ is a dimensionless coupling constant, 
$\phi$ is the Quintessence field, also indicated by Q, 
and $\phi_{0}$ is its value today. 

As a first step, we investigated the cosmological trajectories 
in this scenario, focusing on the redshift evolution of the 
Q energy density. The Q potential is modeled as 
an inverse power law, namely 
$V(\phi )=M^{4+\alpha}/\phi^{\alpha}$, where $\alpha >0$ and 
$M$ is the energy scale chosen to have the amount of Q energy today. 

In the radiation dominated era we find that the divergence of $R$ 
at early times generates an effective gravitational potential 
in the Klein Gordon equation, which drives the Q evolution in a sort 
of gravitational slow rolling. We named this era $R$-boost. 
The dominant component in $\rho_{\phi}$ during this epoch is the kinetic 
component. The $R$-boost ends when this kinetic energy becomes 
comparable with the true potential energy: we have 
estimated that this time corresponds to a redshift $1+z_{pot} \simeq
3000$, i.e. the $R$-boost lasts also for the earliest part of the
matter dominated era. 

In the matter dominated era we obtained scaling solutions 
in NMC theories, which extend the validity of those studied in \cite{Uzan}, 
for which $\rho_{\phi}$ and the Q pressure $p_{\phi}$ scale as follows: 
\begin{equation}
\label{summaryrhoqtrack}
\rho_{\phi}=\rho_{Q0}\left({a_{0}\over a}\right)^{\epsilon}
\ \ ,\ \ \epsilon ={3\alpha\over \alpha + 2}\ \ ,
\ \ p_{\phi}=-\rho_{\phi}{2\over \alpha +2}\ .
\end{equation}
This regime holds for almost all low redshifts,  $1+z\lesssim 100$. 
At recent times, however, $1+ z \lesssim 3$, the Quintessence 
component starts to dominate and the expansion accelerates. 

These results show that the Extended Quintessence is a tracker 
field, in the sense of the original works on Quintessence 
tracker solutions \cite{Ratra,track}. While the Quintessence 
potential energy density has to be chosen to produce the 
observed $\Omega_{\phi}$ today, just as if Q was ordinary 
cosmological constant, the initial amount of $\rho_{\phi}$ 
can vary over a very wide range, covering more than 20 orders 
of magnitude. 

We also investigate perturbation spectra in TEQ. The tracker 
dynamics imprints considerably larger effects 
compared with ordinary Tracking Quintessence and the 
cosmological constant, both on CMB and LSS. 
We gave quite accurate analytical formulas to describe these effects, 
that we summarize here. 

We have shown that in TEQ models the low redshift dynamics 
of the field is either increased or decreased depending on whether $\xi$ is 
positive or negative, respectively. 
This leaves its imprints  on the large scale CMB 
angular power spectrum, where the signal is unprocessed 
by acoustic oscillations, and is affected by the 
dynamics of the gravitational potential (Integrated 
Sachs Wolfe effect, ISW). We derived
the following simple formula describing this effect. 
If $C_{\ell}$ is the coefficient describing the power 
of CMB anisotropies on large angular scales, 
$\ell\lesssim 10$, the shift in TEQ models compared with ordinary Q is
given by 
\begin{equation}
\label{summaryisw}
{\delta C_{\ell\lesssim 10}\over C_{\ell\lesssim 10}}\simeq 
12\left(1-8\pi GF_{dec}\right)\ ,
\end{equation}
where $F$ is the NMC function in Eq.(\ref{summaryf}). 
because of the tracking regime we generally have $\phi_{beg}\ll\phi_{0}$, 
so that the above equation assumes the simple form 
\begin{equation}
\label{summaryiswnumbers}
{\delta C_{\ell\lesssim 10}\over C_{\ell\lesssim 10}}
\simeq 96\pi G\xi\phi_{0}^{2}\ ,
\end{equation}
which  means an effect going 
from $10$ to $50\%$ for the potential slope $\alpha$ increasing from 
$1$ to $3$. 
Note that for negative $\xi$ the ISW is inverted, a feature 
that is quite unusual in Quintessence models. 

The amplitude of the acoustic peaks varies because of the same ISW 
effect, if the overall power is normalized to COBE at $\ell =10$. 
There exists also another mechanism that changes 
the acoustic peaks amplitude, which occurs at decoupling 
rather than at low redshifts. 
The Hubble horizon value at a given redshift is different 
in Extended vs. ordinary Quintessence models, 
as already noticed in \cite{PBM}. This is because in order to reach 
the same value today the Hubble parameter followed a different 
dynamics in the past; we have shown here that 
the horizon reentry for a given 
comoving scale is delayed for positive $\xi$ and 
anticipated for negative $\xi$. This implies a slight 
suppression of the acoustic oscillations in the first case, 
and an enhancement in the second, as a consequence of the 
different amounts of CDM when the baryon and photon
densities are oscillating. 

TEQ imprints also affect the position of 
the acoustic peaks. The angular multipoles at which 
CMB acoustic oscillations are occurring are inversely 
proportional to the comoving size of the CMB sound horizon 
at decoupling, proportional to $\tau_{dec}$,  
and directly proportional to the comoving 
distance between last scattering and us, 
$\tau_{0}-\tau_{dec}$. Both these quantities are affected 
by the TEQ dynamics. The CMB sound horizon is proportional to the Hubble 
horizon at decoupling, which, as we just mentioned 
is different in the various models. 
The comoving distance 
between last scattering and us is also affected by the 
TEQ low-redshift dynamics. 
We derived the following simple analytical formula 
giving the shift in the positions of the acoustic peaks, 
which accurately describes our numerical results: 
\begin{equation}
\label{lshiftconclu}
{\delta\ell\over \ell}\simeq {8\pi G F(\phi_{dec})-1\over 8}=
\pi G\xi\phi_{0}^{2}\ .
\end{equation}
Note that again, depending on the sign of $\xi$, 
Eq.(\ref{lshiftconclu}) may imply a peak shift towards 
smaller or larger angular scales. 

For what concerns the LSS effects today, the first point 
to make is that since at the COBE scales, where we 
normalize our models, 
the CMB spectrum is affected by the ISW, 
the amplitude of  perturbations on LSS scales is suppressed 
if $\xi >0$ and increased if $\xi <0$, by an amount that is 
approximately the opposite of the quantity in 
Eq.(\ref{summaryiswnumbers}). 
Second, the turnover scale position $k_{turn}$ is 
shifted, being inversely 
proportional to the Hubble horizon at the equivalence. 
The size of this effect can be simply estimated by the 
Friedmann equation: 
\begin{equation}
\label{summarypeakshift}
\left(\delta k_{turn}\over k_{turn}\right)_{NMC}=
{1-8\pi G F_{eq}\over 2}\ .
\end{equation}
Again, this is roughly $4\pi G\xi\phi_{0}^{2}$ because, 
thanks to the tracking behavior, $\phi_{eq}\ll\phi_{0}$. 

This completes the list of our main results. 
A very final comment we want to make is that 
CMB anisotropies have been shown here to be considerably affected 
by the dynamical properties of the vacuum energy and by the 
Non-Minimal Coupling terms present in the Lagrangian. 
This is an important novelty in the literature on Quintessence models, which 
deserves as much attention as the dependence of the CMB 
spectra on the kind of dark matter, value of the primordial 
perturbation spectral index, etc..  
This is made possible by the fact that the CMB 
itself is more than a snapshot of the early universe; 
its properties in fact are also determined by a sort of line-of-sight 
integration over a very long part of the cosmic evolution, 
which is then able to tell us about the time variation of the gravitational 
constant and the dynamics of the vacuum energy. 

With higher and higher accuracy achieved in CMB data, going from 
the encouraging, present-day ones \cite{Boom,Max} to the formidable 
performance of the Microwave Anisotropy Probe and Planck mission 
\cite{CMBFUTURE}, we will be able to get exciting new insights 
into the structure of gravity as well as the nature of the 
vacuum energy.

\acknowledgements Matthias Bartelmann, Andrew Liddle 
and Uros Seljak are warmly acknowledged for useful comments.

\end{document}